\documentclass[fleqn,10pt]{wlscirep}
\usepackage[utf8]{inputenc}
\usepackage[T1]{fontenc}
\title{Time Localized Tilted Beams in Nearly-Degenerate Laser Cavities}

\author[1,2]{A. Bartolo}
\author[2]{N. Vigne}
\author[1]{M. Marconi}
\author[3]{G. Beaudoin}
\author[3]{K. Pantzas}
\author[3]{I. Sagnes}
\author[2]{A. Garnache}
\author[1,*]{M. Giudici}

\affil[1]{Université Côte d'Azur, CNRS, Institut de Physique de Nice, 06560 Valbonne, France}
\affil[2]{Institut d'Electronique et des Systèmes, CNRS UMR5214, 34000 Montpellier, France}
\affil[3]{Centre de Nanosciences et de Nanotechnologies, CNRS UMR 9001, Université Paris-Saclay, 91120 Palaiseau, France.}
\affil[*]{massimo.giudici@univ-cotedazur.fr}

\begin{abstract}
We show that nearly-degenerate Vertical External-Cavity Surface-Emitting Lasers may emit a set of tilted beams of individually addressable mode-locked pulses. These time localized beams feature a Gaussian profile and they are emitted in pairs with opposite transverse k-vector. Because they are phase locked, their interference leads to a non homotetic pattern in the near-field emission of the laser. In the simplest situation, when a single pair is emitted, this is a stripe pattern. Our analysis discloses the role of third order (spherical) aberrations of the cavity in stabilizing this spatio-temporal mode-locked regime and in selecting the value of the transverse k-vector.
\end{abstract}

\begin{document}

\flushbottom
\maketitle
\thispagestyle{empty}

\section*{Introduction}

Multimode photonics is a novel research subject  devoted to the generation and control of
complex light states for applications to information processing, photonic computing, sensing,
and imaging \cite{Forbes2021,Wright:22,Piccardo_2022,Davidson:22,SMS-SR-16,CBT-OE-18}. In lasers the customization of the emitted spatial profile requires  a degenerate cavity \cite{Arnaud:69} and a broad-area pump. These large aspect-ratio (or large Fresnel number) laser systems have been implemented using solid-state lasers \cite{Forbes2013,Tradonsky:21} and Vertical External-Cavity Surface-Emitting Lasers (VECSELs) \cite{Knitter:16,Cao2019,Piccardo22}. In large Fresnel number resonators multistability between different emission profiles may lead to Localized Structures (LS) \cite{TML-PRL-94,rosanov,BLP-PRL-97,1172836} which, in their simplest form, are individually addressable bright spots of light in the transverse section of the laser cavity. Spatial LS can be used as fundamental bricks for structuring laser light, as experimentally shown in broad-area Vertical Cavity Surface-Emitting Lasers (VCSELs) \cite{BTB-NAT-02,TAF-PRL-08,GBG-PRL-08,GBG-PRL-10}. Localized structures have been also observed in the longitudinal dimension of optical resonators driven by an injected field: these temporal LS (TLS) are individually addressable pulses circulating inside the cavity \cite{LCK-NAP-10,HBJ-NAP-14,PPR-PR-18,Leo2021}. In semiconductor lasers, TLS  have been implemented within
the regime of passive mode-locking induced by a semiconductor saturable absorber mirror (SESAM)
\cite{MJB-PRL-14,CSV-OL-18}. When the cavity round-trip time is larger
than the gain recovery time and the modulation
depth of the saturable absorber overcomes a critical value \cite{GJ-PRA-17}, a variety of mode-locked states with
a different number of pulses per round trip coexist with the off solution.
In these conditions, mode-locked pulses can be individually addressed \cite{CJM-PRA-16}. 

More recently, multimode lasers in the three spatial dimensions have been demonstrated. Spatio-temporal mode locking has been reported in optical fibers \cite{Wright17,WSP-Nature-20,DXL-PRL-22} and temporally localized Turing patterns have been observed in VECSEL with SESAM \cite{BVM-Optica-22}. 

In this paper we analyze the emission of a nearly-degenerate VECSEL cavity operated in the regime of TLS when the laser cavity is unstable for an axial emission. While this laser platform emits temporally localized patterns with an hexagonal structure in far-field when axial emission is supported \cite{BVM-Optica-22}, here we show that third-order (or spherical) aberrations stabilize a set of Gaussian-like tilted beams whose interference leads to non homotetic temporally localized spatial patterns. This experimental result discloses the leading role of aberrations in nearly-degenerate cavities and it gives evidence of a novel regime of spatio-temporal mode-locking based on phase-locked tilted beams circulating in the cavity.

\section*{Results}

\begin{figure}[h]
    \centering
\includegraphics[width=0.5\columnwidth]{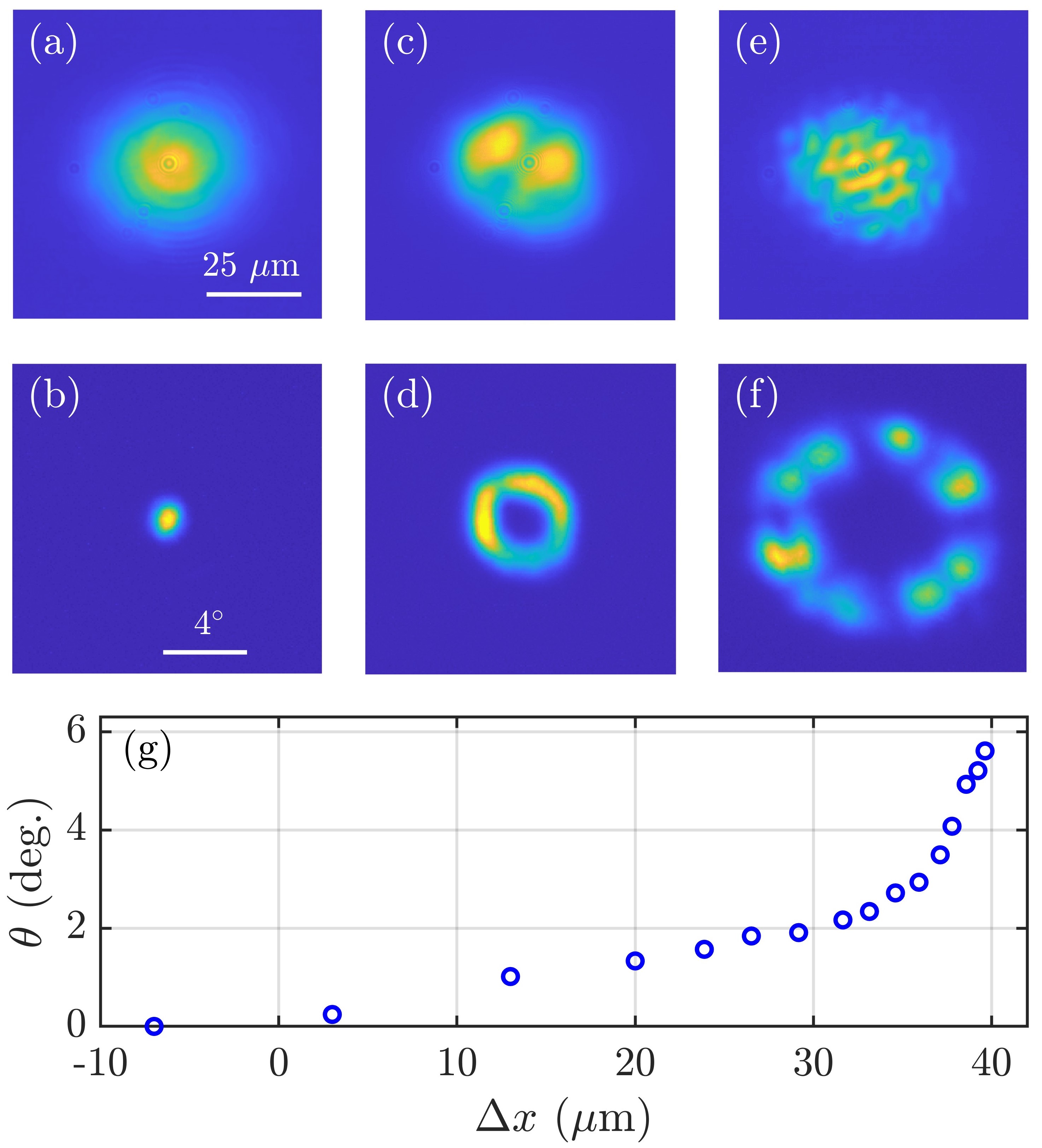}
\caption{The VECSEL cavity ABCD matrix parameters are set to $C\gtrapprox0$ ($\Delta z$=-2.74 mm) and $B\gtrapprox0$ ($\Delta x$ is varied in the range $-50\mu m<\Delta x< 40\mu m$). near-field (a,c,e) and far-field (b, d, f) emission profiles for three positions of the SESAM: (a) and (b) $\Delta x=-7 \mu m$, (c) and (d) $\Delta x=29 \mu m$ and (e) and (f) $\Delta x=38 \mu m$. (g) Tilting angle of the emitted beams with respect to the optical axis as a function of $\Delta x$.}
\label{fig:patterns}
\end{figure}

We consider a large aspect-ratio Vertical External-Cavity Surface-Emitting Laser (VECSEL) operated in the regime of TLS, i.e. featuring i) cavity round-trip time ($\tau$)  larger than the carrier's recombination time ($\tau_g$) and ii) SESAM's saturable losses larger than a critical value (typically $\Delta R>8\%$, \cite{CSV-OL-18}). Large aspect-ratio is achieved by using a nearly self-imaging external cavity and by pumping the gain mirror with a flat-top elliptical beam having a size of $90\times 50 \mu\mathrm{m}$. The details of the experimental setup are described in Methods.

The spatial profile of the VECSEL emission depends on the $B$ and $C$ values of the ABCD ray transfer matrix describing the round-trip propagation of the field in the external cavity. As shown by the Huygens-Fresnel diffraction integral \cite{SIEGMAN-BOOK,theory}, $B$ controls diffraction of the beam travelling in the cavity, while $C$ controls its wavefront curvature. Close to the self-imaging condition (SIC) and for the choice of the optical elements used in our external cavity (see Methods), the value of $B$ can be controlled by shifting the SESAM around its SIC position ($x_0$) while the value of $C$ can be controlled by shifting one of the cavity lenses (cf. $L_2$ in Fig. \ref{fig:setup}a) around its SIC position ($z_0$). By calling these displacements $\Delta x=x-x_0$ and $\Delta z=z-z_0$ respectively, we find $B=8\Delta x$ while $C=-2\,\Delta z/64 \mathrm{mm}^2$ (see Eq.~\ref{eq:1}) \cite{BVM-Optica-22,BVM-Optica-22SuppMat}.

For positive diffraction and for focusing wavefront curvature ($B>0,C<0$), a temporal localized hexagonal pattern appears when approaching SIC and it is emitted together with a Gaussian axial mode \cite{BVM-Optica-22}. 

When approaching SIC for an overall defocusing wavefront curvature ($C>0$), the cavity is stable for an axial emission only for negative diffraction ($B<0$), i.e. $\Delta x<0$. In the experiment, for $-50\mu m<\Delta x<0\mu m$, the VECSEL emits an axial
fundamental Gaussian mode (Fig. \ref{fig:patterns}(a) and (b)) whose waist decreases as $\Delta x$ is
increased. For $\Delta x>0$ no axial mode can be supported for emission. However, instead of switching off,
the laser emits non homotetic patterns having a far-field distribution in the form of a ring (Fig. \ref{fig:patterns}(c) and (d)). The diameter of this ring increases with $\Delta x$ and, correspondingly, the spatial frequency of the near-field profile increases, as shown in Fig. \ref{fig:patterns}(g). The system is emitting a set of tilted plane-waves having transverse wavevectors in different spatial directions but all sharing the same module. The circular symmetry of the far-field emission is rather fragile and it is broken for large value of the transverse wavevectors. Then, the circle in the far-field breaks up in spots \ref{fig:patterns}(f) which are paired: each transverse wavevector $\overrightarrow{k_{\perp}}$ coexists with its opposite one $-\overrightarrow{k_{\perp}}$. These two waves are phase locked and their interference gives birth to fringes onto the surface of the gain section and of the SESAM (as in Fig. \ref{fig:patterns}(c) and (e).

\begin{figure}
\centering
 \includegraphics[width=0.99\columnwidth]{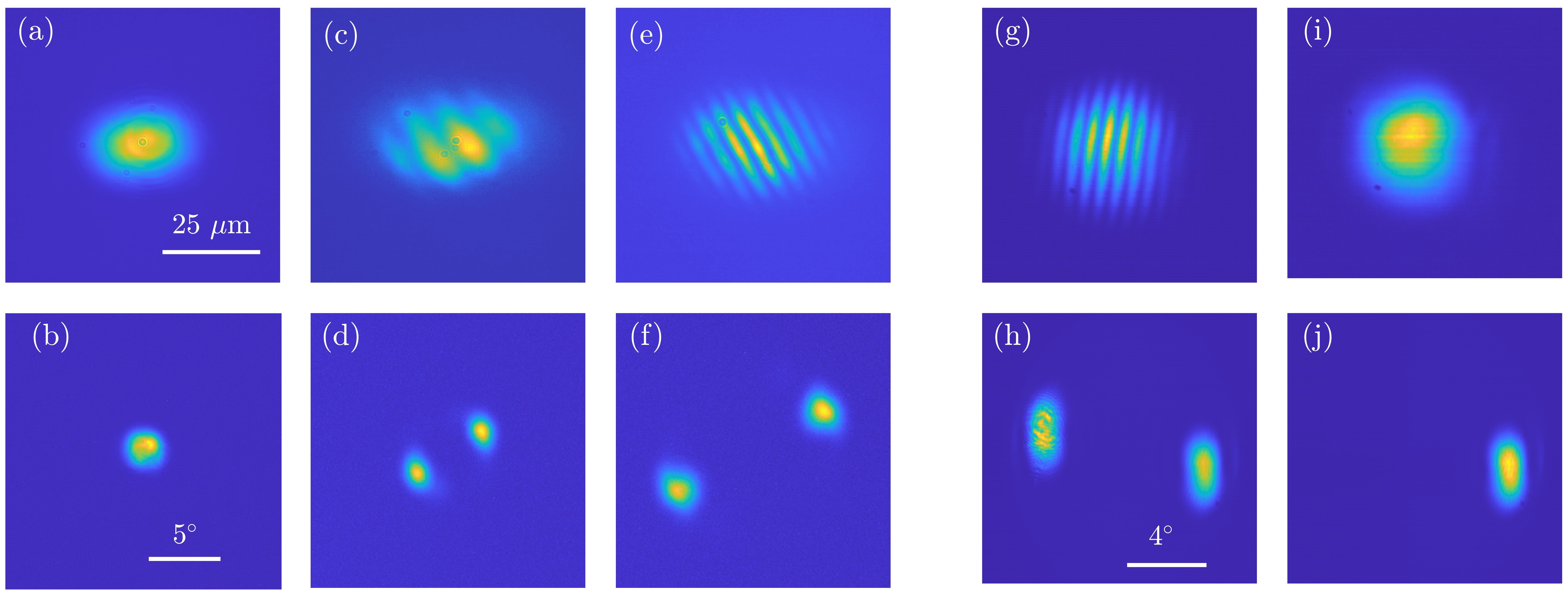}\caption{Near-field (a, c, e) and far-field (b, d, f) profiles of VECSEL emission in presence of a small anisotropy in the laser cavity. for different values of $\Delta x$: (a),(b) $\Delta x$ = 0$\mu m$, (c),(d) $\Delta x$ = 30$\mu m$ and (e),(f) $\Delta x$= 35$\mu m$;  $\Delta z$= -2.74 $mm$ and $P_p$=175$mW$. In panels g) et h) the direction of the anisotropy has been changed and the direction of the pair of tilted beams is different. In (i) and (j) one of the two beams ($-\protect\overrightarrow{k_{\perp}}$)
		has been filtered out by placing a circular aperture in a Fourier
		plane of the detection path (see Panel (j)). Panel (i) show the near-field distribution
		of the remaining tilted beam, revealing a Gaussian profile.
        \label{fig:twospots}}
\end{figure}

Conical emission of tilted waves is broken by inserting an anisotropic element in the set-up (as a glass window) or simply by slightly tilting an optical element of the cavity ($L_2$ $L_3$ or the SESAM). Then, only two points are left in the far-field (Fig. \ref{fig:twospots}), thus revealing the emission of two tilted waves having opposite transverse k-vector $\pm \overrightarrow{k_{\perp}}$. Interference of the two beams gives birth to a stripe pattern onto the surface of the gain section and of the SESAM. The angular dependence of the tilted wave emission as a function of $\Delta x$ follows the same curve shown in Fig. \ref{fig:patterns} (g).

For examining the intensity profile of a single tilted wave circulating in the cavity we may prevent interference between the two opposite wave vectors outside the cavity. This has been done by analyzing the situation of a single pair of tilted beams and filtering one of the two spots in a far-field plane of the detection path. The near-field profile of the single beam remaining exhibits a Gaussian profile, as shown in Fig. \ref{fig:twospots}, panels i) and j).

\begin{figure}
\centering
	\includegraphics[width=0.7\columnwidth]{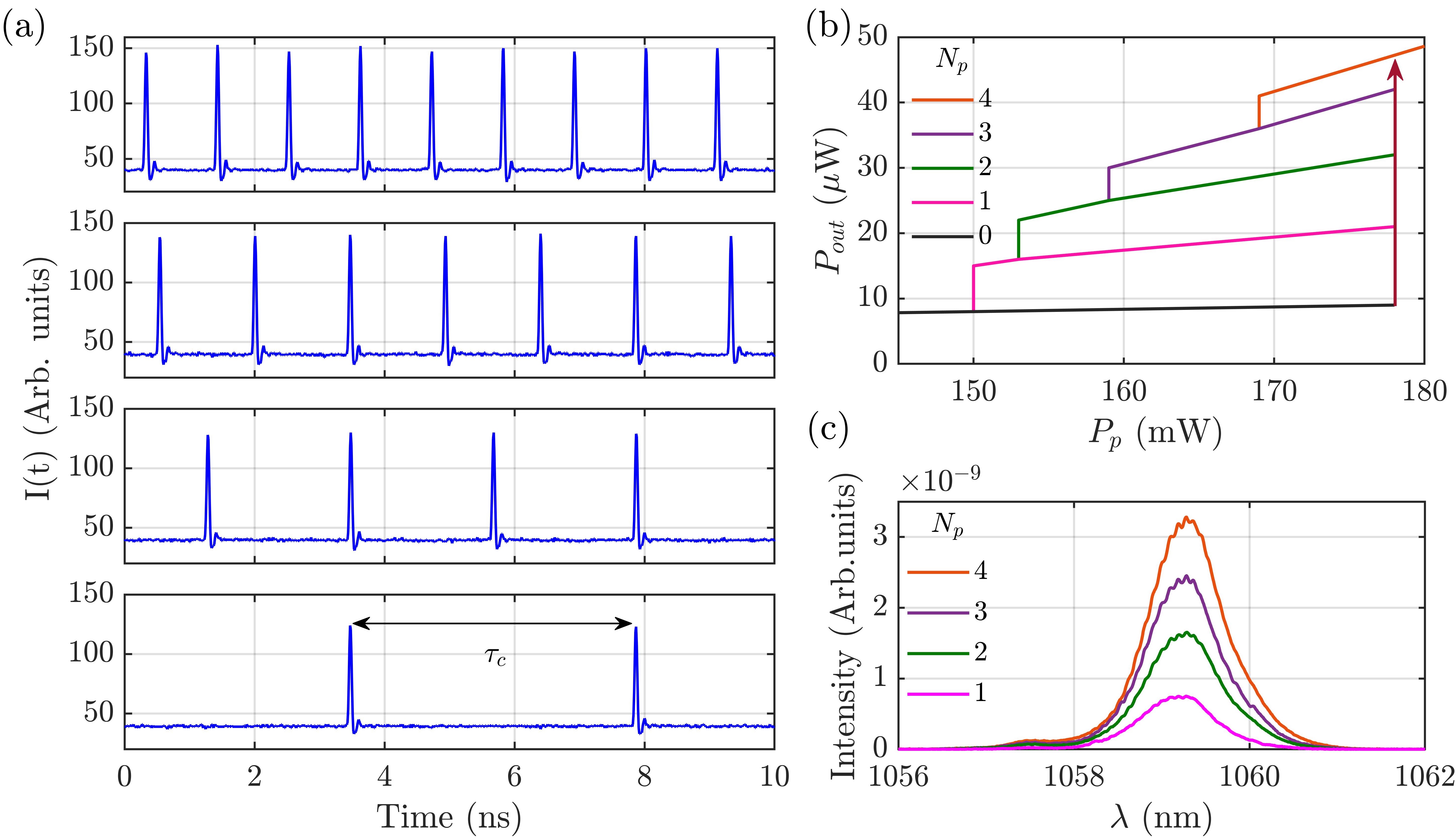}
	\caption{a) Coexisting time resolved output of the VECSEL corresponding to the profile shown in Fig. \ref{fig:twospots} c) and d). These mode-locked states with a number of pulses per roundtrip ranging from one to four correspond to the same pattern profile upon an intensity scaling factor which depends on the number of pulses. The width of the pulses emitted is below the time resolution of our detection system hence their width is smaller than 20 ps.. b) Total output power emitted by the VECSEL versus the pump power $P_{p}$ showing the stability of the different pulsating states and of the off solutions, ranging from no pulse to four pulses
		per roundtrip. c) Optical spectra for each pulsing state shown in a)}\label{temporal}
\end{figure}

The VECSEL emission profiles described above are acquired with CCD camera which cannot resolve the temporal behavior of the laser emission. The analysis of laser dynamics is obtained by using an array of fast detectors capable of monitoring different regions of the near-field emission, see Methods for more details. This analysis shows that the time behavior of the patterns of Fig. \ref{fig:patterns} and \ref{fig:twospots} corresponds to mode-locked states with a number of pulses per roundtrip spanning from one to four, as shown in Fig. \ref{temporal}. These states and the off solution coexist in a large interval of the pump level, as shown in Fig. \ref{temporal}b). Their stability ranges $P_{p,a}<P_p<P_{p,b}$ share the same upper limit $P_{p,b}$=178 mW, which corresponds to the VECSEL threshold, while the lower limit $P_{p,a}$ increases with the number of pulses per round-trip. Coexistence of the four states is observed for 150 mW<$P_p$<178 mW. This multistability is the signature of the TLS regime, where the pulses can be individually addressed by shining short pump pulses onto the VECSEL \cite{CSV-OL-18,BVM-Optica-22}. It is worth to underline that these different states are associated to the same near- and far-field profile upon an intensity scaling factor which depends on the the number of pulses travelling in the cavity. 
 In Fig. \ref{temporal} c) we show also the optical spectra for each pulsing state shown in Fig. \ref{temporal} a), they disclose a spectral envelope of $\approx 1$ nm, FWHM. It is worth noting that the resolution of the optical spectrum analyzer (0.06 nm, i.e. 15 GHz) does not allow to resolve the spectral combs corresponding to the mode-locked states shown in Fig. \ref{temporal} a) whose teeth separation is in the range of .25-1.25 GHz.  Finally, we precise that the spatially resolved detection indicates that the pulsing activity is synchronous throughout the entire near-field section of the pattern emitted by the VECSEL.

\begin{figure}[h]
    \centering
\includegraphics[width=0.5\columnwidth]{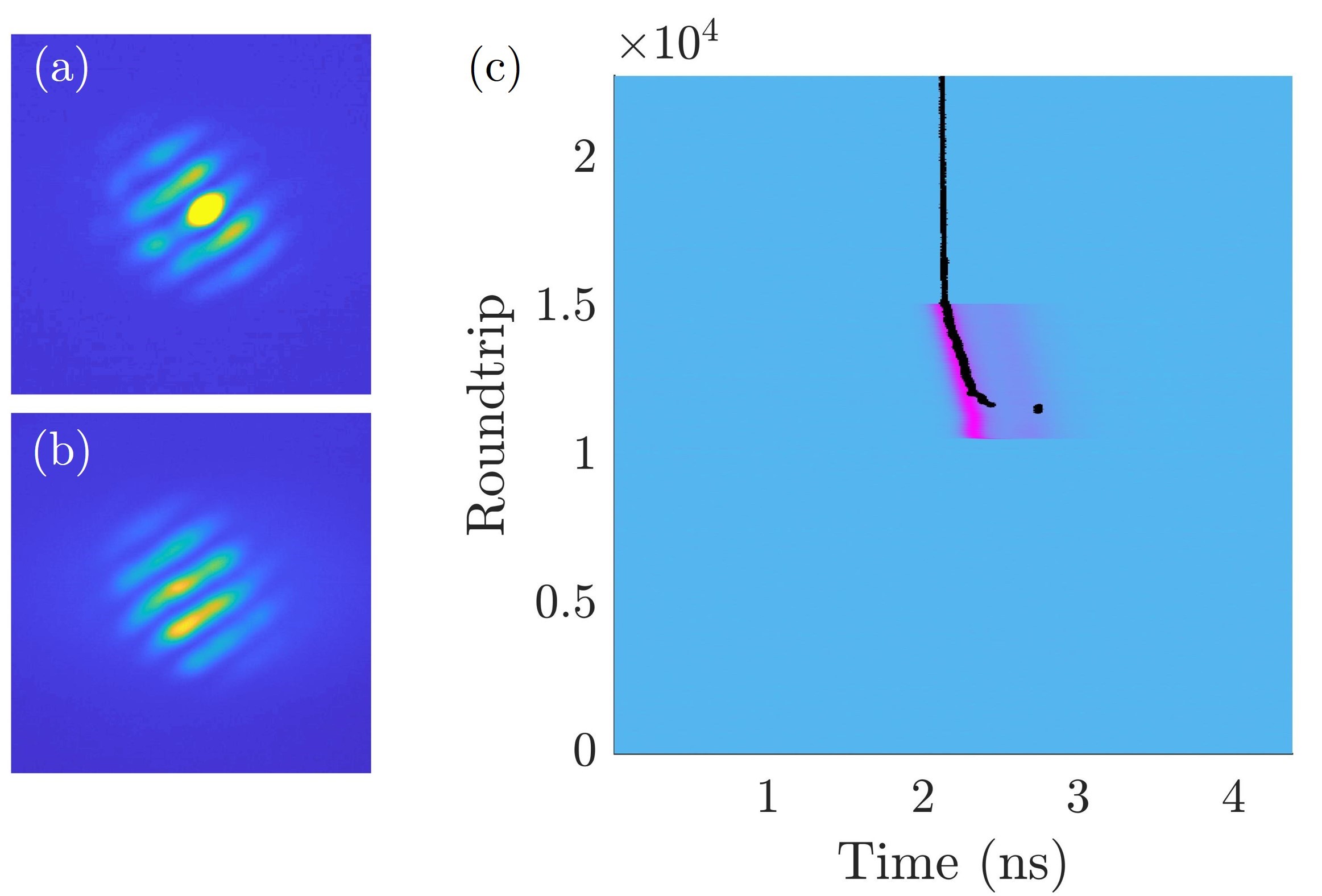}
\caption{(a) Near-field profile of the stripe pattern when the perturbation beam targeting one stripe of the pattern is applied (a) and after removal of the perturbation beam (b). The position in the gain mirror section targeted by the addressing beam is important for a successful addressing. One must target the region where, at
the switching, a bright fringe will appear in the near-field, as shown in panel a). (c) Spatiotemporal diagram of the writing process of the temporal localized
stripe pattern. The evolution of the TLS roundtrip after rountrip is represented by a black trace, while the pump
evolution is represented by using a color code. 
Stationary value of the pump is  = 174 mW and addressing is obtained by sending
a 120 ps pump pulse, 10 mW peak power onto the gain mirror between round-trip number 10800 and round-trip number 15400. }
\label{fig:switchon}
\end{figure}

In our optically pumped VECSEL temporal LSs can be individually addressed by shining short pump pulses onto the gain mirror. The system is set by the CW optical pump in the multistable parameter region ($150$mW$<P_{p}<178$mW) 
where LSs exist. An additional laser, capable of providing optical pulses of about 120 ps (FWHM) at arbitrary rate, is used to generate pump pulses which are overlapped to the CW pump. These pulses have a Gaussian spatial profile and a waist of
8 $\mu m$ which matches approximately the size of the stripes of the emitted pattern (Fig. \ref{fig:switchon}a). Pulses peak power is
chosen to be sufficiently large to drive the VECSEL beyond the upper
limit of the multistable region shown in Fig. \ref{temporal}b), where only the solution composed
by four pulses per roundtrip is stable.  Finally, the addressing pulse is sent
to the gain mirror synchronously with the cavity roundtrip for about
five thousand roundtrips. The addressing process is depicted in Fig.~\ref{fig:switchon}
by using a space-time diagram where the pump value is represented by
using a color code, while the trajectory of the LS is represented
by a black trace.  In Fig. \ref{fig:switchon}, we choose an initial
condition where no LS is present inside the cavity before the addressing
pulse. The pump pulse is sufficiently short to switch-on a single
LS which persists after the perturbation is removed.

Other initial conditions can be chosen with similar results, provided that the addressing pulse is
separated in time from the preexisting LS of at least $\tau_{g}$. This addressing technique can be adapted for erasing LSs \cite{BVM-Optica-22}. Erasure of LSs can be also obtained by feedbacking an emitted pulse onto the gain mirror with proper time delay, as described in \cite{Bartolo:21}.

\section*{Discussion}

Continuous wave tilted beam emission resulting in stationary Turing patterns have been previously observed in large aspect-ratio laser \cite{FMN-PRL-93,Weiss,PhysRevLett.82.1434}. In these situations a positive detuning between the laser gain resonance
and the closest resonator resonance is at the origin of this instability. The laser emits tilted beams whose frequency matches the gain peak and whose wavevector exhibits a longitudinal component matching the cavity resonance. This mechanism does not apply to our system where the set of longitudinal
cavity resonances is very dense (less than 500 MHz free spectral range)
compared to the width of all other relevant spectral filtering curves,
such as microcavities resonances (> 9 nm, i.e. more than 1 THz), gain and saturable absorption curves (more than 10 THz). Moreover, we have verified that the angle of the tilted waves emitted is not depending on any parameter of the active or passive media (temperature and pumping). As shown in Fig. \ref{fig:patterns}g), this angle depends instead on the distance between the SESAM and its collimator L$_4$. 

In our system, the origin  of tilted waves can be understood by considering spherical - or third order- aberrations of the lenses used in the cavity. The most important contribution to aberrations comes from the short focal length collimators and, in particular, from $L_4$ where, because of the magnification factor $M$ (see Methods), the incident beam is at the largest distance from the optical axis. The effect of aberrations on the beam path is schematically shown in Fig. \ref{fig:setup}, where we assume a dependence of the focal length of $L_4$ on the angle $\theta=\rho/f_c$ of the incident beam, $f(\theta)=f_0(1+\alpha \theta ^2)$ \cite{Yoshida:82,Aruga:97}, being $\rho$ the radial distance between the optical axis and the incidence point of the incoming beam. Accordingly, if $\alpha>0$, for a range of distances $d_5$ larger than $f_0$ there is a set of off-axis  Gaussian beams which is stable, as shown schematically in figure \ref{fig:setup} b). By using a perturbative approach \cite{nathan}, we have calculated the impact of spherical aberration on ABCD roundtrip matrix. This enables to determine the stability of axial and tilted Gaussian beams in the cavity as a function of the cavity parameters $B$ and $C$. The tilting angle and the waist of the Gaussian beam allowed in the resonator is plotted in fig. \ref{fig:setup}c), which agrees qualitatively with the experimental observations for small values of $\Delta x$ ($\Delta x<30 \mu m$). For larger value of $\Delta x$ the perturbative approach fails and higher order corrections should be taken into account. 

It is worth to point out that the sign of $\alpha$ depends mainly on $L_4$ parameters and, by using other lenses, it was possible to test the effects of negative values of spherical aberration $\alpha$. In these cases, we observe the tilted beam angle increasing by shortening the distance $d_5$, thus confirming our interpretation \cite{nathan}. It is worthwhile noting, that when the SESAM is replaced by a high reflective mirror, stationary tilted beams are emitted by the VECSEL \cite{Hachair:08}. In this situation, they do not necessarily appear in pairs and a single transverse wavevector tilted beam can be observed in the cavity \cite{nathan}. This observation illustrates that the emission of pairs of phase locked tilted beams is imposed by the presence of a saturable absorber. Interference between the paired beams leads to intensity modulation onto the SESAM surface which saturate the absorber efficiently and enables laser emission.   

 In conclusion, we have analyzed a nearly-degenerate VECSEL cavities in the regime of TLS and we have given evidence of a novel spatio-temporal mode-locking regime supported by spherical aberrations of the laser resonator. A phenomenological description of third order aberrations by using the perturbed ABCD roundtrip matrix \cite{nathan}, explains how spherical aberrations stabilize tilted Gaussian beams in nearly-degenerate cavities (see Methods). Our experimental observations are in good qualitative agreement with a recently published theoretical model \cite{theory}. Possible applications are related to the ability of addressing individually the pulses circulating in the cavity at different azimuthal angles with respect to the optical axis. These pairs of beams can be independently controlled at any far-field plane inside the cavity, which leads to the possibility of customizing the spatio-temporal structure of the emitted profile. This control can be useful for information processing, photonic computing and for generating multiple frequency comb beams \cite{Keller22}.

\section*{Methods}

\begin{figure}
\centering
	\includegraphics[width=0.9\columnwidth]{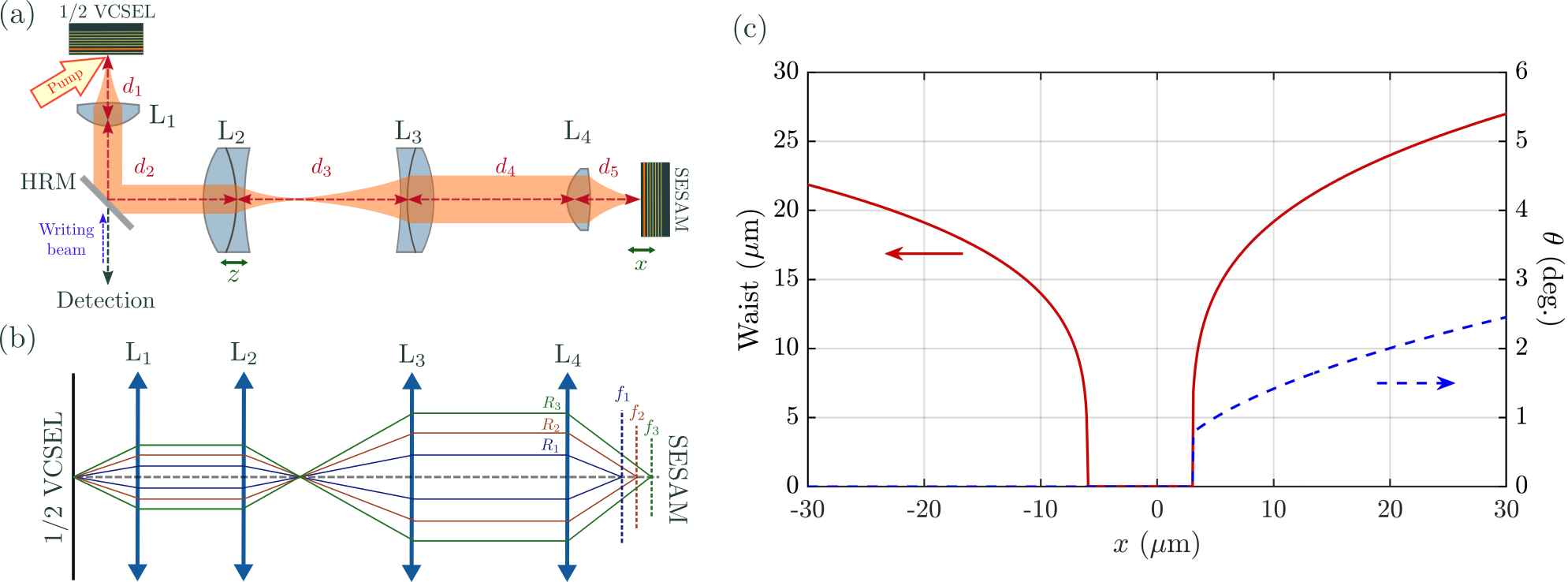}\caption{Panel a) Experimental set-up showing the L-shape nearly-8f cavity VECSEL. $d_{1}$:
		distance between the gain section and lens $L_{1}$, $d_{2}$: distance
		between $L_{1}$ and lens $L_{2}$, $d_{3}$: distance between $L_{2}$
		and lens $L_{3}$, $d_{4}$: distance between $L_{3}$ and lens $L_{4}$,
		$d_{5}$: distance between $L_{4}$ and the SESAM, HRM= high reflectivity
		beam splitter (>99.5$\%$ at 1.060 nm). Panel b) Simplified scheme of the path of pairs of tilted beams travelling at different angles in a 4-lenses telecentric cavity. Only $L_4$ is supposed to exhibit spherical aberrations with $\alpha$>0. Panel c) Stability of the Gaussian beam circulating in the cavity. red curve: radius of the Gaussian beam at $1/e^2$ on the gain section,  blue curve:  angle of tilting (degrees) with respect to the optical axis of the stable Gaussian beam.  \label{fig:setup}}
\end{figure}

The VECSEL cavity has an L-shape and it is delimited by a gain mirror (also called 1/2 VCSEL) and by a semiconductor SA mirror (SESAM), as shown in Fig. \ref{fig:setup}aa). The former is based on a GaAs substrate with 12 strain-balanced InGaAs/GaAsP quantum wells (QWs) designed for barrier optical pumping and emitting at 1.06 $\mu\mathrm{m}$. The gain mirror is optically pumped at $808\,$ nm by a flat-top elliptical profile having an horizontal
axis of 90 $\mu$m and a vertical one of 50 $\mu$m. The SESAM features a single strained InGaAs/GaAs QW located at $1\sim2$~nm from the external surface leading to a carrier's recombination rate approximately two orders of magnitude faster than the gain medium which is compatible with
a stable passive mode-locking and with the existence of temporal localized states
\cite{SJG-PRA-18}. Light extraction from the cavity
occurs by transmission through a high reflective beam splitter (>99.5
reflectivity at 1.060 nm). The output beam from the VECSEL is sent to the detection part
where the far-field and near-field profiles are imaged on two CCD
cameras. The near-field is also imaged on an array of optical fibers
for spatially resolved detection at 10 GHz bandwidth. Finally, the
total emission is monitored by a 33 GHz bandwidth detection system
and by an optical spectrum analyzer.

Both the gain mirror and the SESAM have been engineered to be operated in a nearly degenerate cavity. The high level of losses of such a cavity has required to increase the confinement factor in the gain mirror by enhancing the microcavity effect. Moreover, for enabling multistable mode-locked states, it is required the SESAM modulation depth of the saturated/unsaturated
reflectivity (also called saturable losses) to be larger than a critical amount ($A=8\%$) \cite{MJB-PRL-14,J-PRL-16,SJG-PRA-18}. By increasing
the finesse of the SESAM microcavity, we achieve a modulation depth $A=23\%$ at the micro-cavity
resonance peak (1.0667 $\mu m$). Fine control of the modulation depth experienced by the intracavity
field is set by controlling the detuning between the two microcavities
cavities ($\delta\lambda=\lambda_{B}-\lambda_{G}$) \cite{CSV-OL-18},
the smaller the detuning the higher the contrast. The results described here has been
obtained for $5.5$ nm$>\delta\lambda>4\,$nm which enables wide multistable
response and a threshold value accessible to our pump power ($P<350$
mW). More details on the samples used can be found in \cite{BVM-Optica-22SuppMat}.

The VECSEL external cavity has been designed to fulfill the requirement
$\tau>\tau_{g}\sim1$~ns and self-imaging condition (SIC) after one
roundtrip. In addition, the SESAM and gain mirror need to be placed
in conjugate planes with a magnification factor $M$ larger than one
for saturating efficiently the SESAM.

Accordingly we use a four-lenses arrangement where the first lens
($L_{1}$), the one closest to the gain section) and the last lens
($L_{4})$, the one closest to the SESAM) are large numerical aperture
collimators (Thorlabs C240TME-1064, $f_{1}=f_{4}=f_{c}=$8 mm) and $L_{2}$ and $L_{3}$
are achromatic lenses (Thorlabs AC254-XX-C) having $f_{2}=100$mm and $f_{3}=$200 mm. Self-imaging condition can be achieved through a telecentric
arrangement of these optical elements, i.e. lenses are placed at distances
given by the sum of their focal lengths ($d_{1}$=$f_{1}$, $d_{2}$=$f_{1}$+$f_{2}$,
$d_{3}$=$f_{2}$+$f_{3}$, $d_{4}$=$f_{3}$+$f_{4}$, $d_{5}$=$f_{4})$. 
thus making a total cavity length $L$=632 mm (cavity round-trip time
$\tau\approx4.2\,$ns) with $M=f_{3}/f_{2}=2$. However, the presence of a pump induced lens onto the gain section
having a focal length $f_{th}$ spanning from 10 to 80 mm depending on the pump level \cite{LMB-OE-09}, requires slight shift of the position of $L_{2}$ and of the position of the SESAM to achieve SIC \cite{BVM-Optica-22SuppMat}.

By calling $x$ the offset of $d_{5}$
with respect to telecentric position ($x=d_{5}-f_{c}$) and $z$ the
offset of $d_{2}$ with respect to telecentric position ($z=d_{2}$$-(f_{c}+f_{2})$) (cf. Fig.~\ref{fig:setup}a),
the SIC in presence of the pump induced lens is obtained at: $z_{0}(f_{th}$)$=-f_{c}^{2}/2f_{th}$
and $x_{0}(f_{th})=-(f_{c}^{4})/(2M^{2}f_{2}^{2}f_{th})$ \cite{BVM-Optica-22SuppMat}. For
focal lengths values in our experiment, one finds that $z_{0}$ is
of the order of few millimeters, while $x_{0}$ will be of few microns
since $f_{2}>>f_{c}$ . 

It is useful to describe the behavior of the $ABCD$ rondtrip matrix close to SIC. By calling $\Delta x=x-x_{0}$ and $\Delta z=z-z_{0}$ the deviations of the SESAM position and of $L_2$ position from the SIC condition, this matrix 
can be approximated to \cite{BVM-Optica-22,BVM-Optica-22SuppMat}

\begin{equation}
	\left(\begin{array}{cc}
		A_{RT} & B_{RT}\\
		C_{RT} & D_{RT}
	\end{array}\right)=\left(\begin{array}{cc}
		1 & \;\;2\,M^{2}\Delta x\\
		-2\dfrac{\Delta z}{f_{c}^{2}}\;\; & 1
	\end{array}\right)\label{eq:1}
\end{equation}
Accordingly, we can identify as $\Delta x$ the experimental parameter controlling
the diffraction coefficient and as $\Delta z$ the experimental parameter
controlling the wavefront curvature. Stability of the cavity requires
$\Delta x>0$ when $\Delta z>0$ and $\Delta x<0$ when $\Delta z<0$.
Third order aberrations, which result in a correction of the vergence for the lenses in the cavity \cite{Yoshida:82,Aruga:97}, can be introduced phenomenologically in the ABCD roundtrip matrix around SIC condition (Eq. \ref{eq:1}), following the perturbative approach developed in \cite{SIEGMAN-BOOK,nathan}. By calling $\delta C(\theta)$ the vergence variation induced by aberrations, $\delta C$ is proportional to the Seidel aberration coefficient $S$ of $L_4$: $\delta C/\theta^2 =S \approx 0.3 mm^{-1}$. Because $|{f_c \delta C}| << 1$, Eq. \ref{eq:1} becomes

\begin{equation}
	\left(\begin{array}{cc}
		A_{RT} & B_{RT}\\
		C_{RT} & D_{RT}
	\end{array}\right)=\left(\begin{array}{cc}
		1 & \;\;2\,M^{2}(\Delta x - f_c^{2}\delta C (\theta))\\
		-2\dfrac{\Delta z}{f_{c}^{2}}\;\; & 1
	\end{array}\right)\label{eq:2}
\end{equation}
This can be used to calculate the stability of the fundamental Hermite-Gaussian beam and its waist in presence of third order aberrations. In Fig. \ref{fig:setup} panel c) we plot the radius and the angle of tilting of the stable beam circulating in the cavity. While for negative diffraction ($\Delta x<0$,  $B<0$), axial Gaussian beams are supported in the cavity, for positive diffraction ($\Delta x >0$, $B>0$), the stable emission is a tilted Gaussian beam whose angle increases with $\Delta x$. There is a qualitative agreement with the experimental curve (Fig. \ref{fig:patterns}g) up to $\Delta x=30 \mu m$, for larger values higher order terms needs to be considered and the perturbative approach described above is not valid anymore.

\bibliography{full_new}

\section*{Acknowledgements }

We acknowledge J. Javaloyes, F. Maucher and S. Gurevich for useful discussions.
We acknowledge funds from Agence National de la Recherche (ANR): projet BLASON (ANR-18-CE24-0002) and projet KOGIT (ANR-22-CE92-0009) and funding from
Région PACA OPTIMAL.

\section*{Author contributions statement}

A.G. and M.G. conceived the experiment(s),  A.B., N.V. and M.M. conducted the experiment(s), A.B., N.V., M.M, A.G. and M.G. analysed the results. G.B., K.P and I.S. realized the semiconductor samples. M.G. and A.G. reviewed the manuscript. 

\section*{Additional information}

\section*{Competing financial interests}
The authors declare no competing financial interests.

\end{document}